\documentstyle[twocolumn,aps,prl,floats,psfig]{revtex}
\begin{document}
\draft 

\title{The Transverse Asymmetry 
$\bf A_{\bf T'}$ from Quasi-elastic $^3\vec{\rm He}(\vec{e},e')$
Process \\and
the Neutron Magnetic Form Factor}

\author{W.~Xu,$^{12}$ D.~Dutta,$^{12}$ F.~Xiong,$^{12}$ B.~Anderson,$^{10}$ 
L.~Auberbach,$^{19}$ 
T.~Averett,$^{3}$ W.~Bertozzi,$^{12}$ T.~Black,$^{12}$ J.~Calarco,$^{22}$ 
L.~Cardman,$^{20}$ G.~D.~Cates,$^{15}$ Z.~W.~Chai,$^{12}$
J.~P.~Chen,$^{20}$ S.~Choi,$^{19}$ E.~Chudakov,$^{20}$ 
S.~Churchwell,$^{4}$ G.~S.~Corrado,$^{15}$ C.~Crawford,$^{12}$ 
D.~Dale,$^{21}$ 
A.~Deur,$^{11,20}$ P.~Djawotho,$^{3}$ B.~W.~Filippone,$^{1}$ 
J.~M.~Finn,$^{3}$ 
H.~Gao,$^{12}$ R.~Gilman,$^{17,20}$ A.~V.~Glamazdin,$^{9}$ 
C.~Glashausser,$^{17}$ 
W.~Gl\"{o}ckle,$^{16}$ J.~Golak,$^{8}$
J.~Gomez,$^{20}$ V.~G.~Gorbenko,$^{9}$ J.-O.~Hansen,$^{20}$ 
F.~W.~Hersman,$^{22}$ D.~W.~Higinbotham,$^{24}$ R.~Holmes,$^{18}$ 
C.~R.~Howell,$^{4}$ E.~Hughes,$^{1}$ B.~Humensky,$^{15}$ S.~Incerti,$^{19}$ 
C.W.~de Jager,$^{20}$ J.~S.~Jensen,$^{1}$ X.~Jiang,$^{17}$ 
C.~E.~Jones,$^{1}$ M.~Jones,$^{3}$ R.~Kahl,$^{18}$ H.~Kamada,$^{16}$ 
A. Kievsky,$^{5}$ I.~Kominis,$^{15}$ W.~Korsch,$^{21}$
K.~Kramer,$^{3}$ G.~Kumbartzki,$^{17}$ M.~Kuss,$^{20}$ 
E.~Lakuriqi,$^{19}$ M.~Liang,$^{20}$ N.~Liyanage,$^{20}$ J.~LeRose,$^{20}$ 
S.~Malov,$^{17}$ D.J.~Margaziotis,$^{2}$ J.~W.~Martin,$^{12}$ K.~McCormick,
$^{12}$
R.~D.~McKeown,$^{1}$ K.~McIlhany,$^{12}$ Z.-E.~Meziani,$^{19}$
R.~Michaels,$^{20}$ G.~W.~Miller,$^{15}$ J.~Mitchell,$^{20}$ S.~Nanda,$^{20}$ 
E.~Pace,$^{7,23}$ T.~Pavlin,$^{1}$ G.~G.~Petratos,$^{10}$ 
R.~I.~Pomatsalyuk,$^{9}$
D.~Pripstein,$^{1}$ D.~Prout,$^{10}$ R.~D.~Ransome,$^{17}$ Y.~Roblin,$^{11}$ 
M.~Rvachev,$^{12}$ A.~Saha,$^{20}$ G.~Salm\`{e},$^{6}$ M.~Schnee,$^{19}$ 
T.~Shin,$^{12}$ 
K.~Slifer,$^{19}$ P.~A.~Souder,$^{18}$ S.~Strauch,$^{17}$ R.~Suleiman,$^{10}$ 
M.~Sutter,$^{12}$ B.~Tipton,$^{12}$ L.~Todor,$^{14}$ M.~Viviani,$^{5}$ 
B.~Vlahovic,$^{13,20}$ 
J.~Watson,$^{10}$ C.~F.~Williamson,$^{10}$ H.~Wita{\l}a,$^{8}$ 
B.~Wojtsekhowski,$^{20}$ J.~Yeh,$^{18}$ P.~\.{Z}o{\l}nierczuk$^{21}$}
\address{$^{1}$California Institute of Technology, Pasadena, CA 91125, USA\\
$^{2}$California State University, Los Angeles, Los Angeles, CA 90032, USA\\
$^{3}$College of William and Mary, Williamsburg, VA~23187, USA\\
$^{4}$Duke University, Durham, NC~27708, USA\\
$^{5}$INFN, Sezione di Pisa, Pisa, Italy \\
$^{6}$INFN, Sezione di Roma, I-00185 Rome, Italy \\
$^{7}$INFN, Sezione Tor Vergata, I-00133 Rome, Italy \\
$^{8}$Institute of Physics, Jagellonian University, PL-30059 Cracow, Poland\\ 
$^{9}$Kharkov Institute of Physics and Technology, Kharkov 310108, Ukraine\\
$^{10}$Kent State University, Kent, OH~44242, USA\\
$^{11}$LPC, Universit\'{e} Blaise Pascal, F-63177 Aubi\`{e}re, France\\
$^{12}$Massachusetts Institute of Technology, Cambridge, MA~02139, USA\\
$^{13}$North Carolina Central University, Durham, NC~27707, USA\\
$^{14}$Old Dominion University, Norfolk, VA~23508, USA\\
$^{15}$Princeton University, Princeton, NJ~08544, USA\\
$^{16}$Ruhr-Universit\"{a}t Bochum, D-44780 Bochum, Germany\\
$^{17}$Rutgers University, Piscataway, NJ~08855, USA\\
$^{18}$Syracuse University, Syracuse, NY~13244, USA\\
$^{19}$Temple University, Philadelphia, PA~19122, USA\\
$^{20}$Thomas Jefferson National Accelerator Facility, Newport News, 
VA 23606, USA\\
$^{21}$University of Kentucky, Lexington, KY~40506, USA\\
$^{22}$University of New Hampshire, Durham, NH~03824, USA\\
$^{23}$Dipartimento di Fisica, Universit\`a di Roma "Tor Vergata", Rome, Italy\\
$^{24}$University of Virginia, Charlottesville, VA~22903, USA}

\date{26 July 2000}

\maketitle

\begin{abstract}
We have measured the transverse asymmetry $A_{T'}$ in
$^3\vec{\rm He}(\vec{e},e')$ quasi-elastic scattering in Hall A at
Jefferson Lab with high
statistical and systematic precision for $Q^2$-values from 
0.1 to 0.6 (GeV/c)$^2$.
The neutron magnetic form factor $G_M^n$
was extracted based on Faddeev calculations 
for $Q^2 = 0.1$ and 0.2 (GeV/c)$^2$ with an experimental uncertainty
of less than 2\%. 
\end{abstract}



The electromagnetic form factors of the nucleon have been a
longstanding subject of interest in nuclear and particle physics.
They describe the distribution of charge and magnetization within
nucleons and allow sensitive tests of nucleon models based on
Quantum Chromodynamics. This advances our knowledge of
nucleon structure and provides a basis for the understanding of more
complex strongly interacting matter in terms of quark and gluon
degrees of freedom.

The proton electromagnetic form factors have been determined with
good precision at low values of the squared four-momentum transfer, $Q^2$,
using Rosenbluth separation of elastic
electron-proton cross sections, and more recently at higher $Q^2$  
using a polarization transfer technique~\cite{jones}.  
The corresponding neutron
form factors are known with much poorer precision because of the lack of free
neutron targets.
Over the past
decade, with the advent of new experimental techniques such as
polarized beams and targets, the precise determination of both the
neutron electric form factor, $G_E^n$, and the magnetic form factor,
$G_M^n$, has therefore become a focus of experimental activity.  Considerable
attention has been devoted to the precise measurement of $G_M^n$.
While knowledge of $G_M^n$ is interesting in itself, it is also
required for the determination of $G_E^n$, which is usually measured
via the ratio $G_E^n/G_M^n$.  Further, precise data for the nucleon
electromagnetic form factors are essential for the analysis of parity
violation experiments~\cite{sample,happex} designed to probe the strangeness content of the
nucleon, where the quantities of interest appear in combination
with the electromagnetic form factors.

Until recently, most data on $G_M^n$ had been deduced from elastic and
quasi-elastic electron-deuteron scattering.  For inclusive
measurements, this procedure requires the separation of the longitudinal 
and transverse cross sections and the subsequent subtraction of a large
proton contribution. Thus, it suffers from large theoretical uncertainties
due in part to the deuteron model employed and in part to 
corrections for final-state
interactions (FSI) and meson-exchange currents (MEC).  The proton
subtraction can be avoided by measuring the neutron in coincidence
($d(e,e'n)$) \cite{Mark93}, and the sensitivity to nuclear structure
can be greatly reduced by taking the cross-section ratio of 
$d(e,e'n)$ to $d(e,e'p)$
at quasi-elastic kinematics.  Several recent
experiments \cite{Ankl94,Brui95,Ankl98} have employed the latter
technique to extract $G_M^n$ with uncertainties of $<$2\%~\cite{Ankl98}
for $Q^2$-values from 0.1 to 0.8 (GeV/c)$^2$.  
While this
precision is very good, there is considerable disagreement among the 
results~\cite{Mark93,Ankl94,Brui95,Ankl98} with respect to the absolute 
value of $G_M^n$. All these exclusive experiments 
require the absolute calibration of the neutron detection efficiency, which
is difficult.

An alternative approach to a precision measurement of $G_M^n$ is
through the
inclusive quasi-elastic reaction $^3\vec{\rm He}(\vec{e},e')$.  In 
comparison to
deuterium experiments, this technique employs a different target and
relies on polarization degrees of freedom.  It is thus subject to
completely different systematics.  A pilot experiment using this
technique was carried out at MIT-Bates 
and a result for $G_M^n$ was extracted \cite{Gao94}.
In this Letter, we report the first precision measurement of $G_M^n$ using
a polarized $^3{\rm He}$ target. 

Polarized $^3$He is useful for studying the neutron
electromagnetic form factors because of the unique spin structure of
the $^3$He ground state, which is dominated by a spatially symmetric $S$
wave in which the proton spins cancel and the spin of the $^3$He
nucleus is carried by the unpaired neutron~\cite{BW84,frier90}. 
The spin-dependent contribution to
the $^3\vec{\rm He}(\vec{e},e')$ cross section is completely contained in 
two nuclear response 
functions, a transverse response $R_{T'}$ and a
longitudinal-transverse response $R_{TL'}$.
 These appear in addition to the
 spin-independent longitudinal and transverse responses $R_{L}$ and
$R_{T}$. $R_{T'}$ and $R_{TL'}$ 
can be isolated experimentally by forming the spin-dependent asymmetry $A$ 
defined as
    $A = (\sigma^{h+}-\sigma^{h-})/(\sigma^{h+}+\sigma^{h-})$,
where $\sigma^{h^{\pm}}$ denotes the cross section for the two 
different helicities of the polarized electrons.
In terms of the nuclear response functions, $A$ can be 
written~\cite{twd86} 
\begin{equation}
\label{asym}
A = \frac{-(\cos{\theta^{*}}\nu_{T'}R_{T'} +
  2\sin{\theta^{*}}\cos{\phi^{*}}\nu_{TL'}R_{TL'})}{\nu_{L}R_{L} +
  \nu_{T}R_{T}}
\end{equation}  
where the $\nu_{k}$ are kinematic factors and $\theta^{*}$ and
$\phi^{*}$ are the polar and azimuthal angles of target spin with
respect to the 3-momentum transfer vector ${\bf q}$. The response functions
$R_{k}$ depend on $Q^{2}$ and the electron energy 
transfer $\omega$. 
By choosing $\theta^\ast = 0$, {\it i.e.} by orienting the target
spin parallel to the momentum transfer ${\bf q}$, one selects the
transverse asymmetry $A_{T'}$ (proportional to $R_{T'}$).

Because the $^3$He nuclear spin is carried mainly by the neutron,
$R_{T'}$ at quasi-elastic kinematics contains a dominant neutron
contribution and is essentially proportional to
${(G_M^n)}^2$, similar to elastic scattering from a free neutron.
Unlike the free neutron case, however, the unpolarized part of the
cross section (the denominator in Eq.\ (\ref{asym})) contains
contributions from both the protons and the neutron in the nucleus.
Therefore, $A_{T'}$ is expected to first order to have the form
${(G_M^n)}^2/(a+b{(G_M^n)}^2)$ in the plane-wave impulse 
approximation (PWIA),
where $a$ is much larger than
$b{(G_M^n)}^2$ at low $Q^2$.  While
measurements of $G_M^n$ using deuterium targets enhance the
sensitivity to the neutron form factor by detecting the neutron in
coincidence, a similar enhancement occurs in inclusive scattering from
polarized $^3$He because of the cancellation of the proton spins in
the ground state.
This picture has been confirmed by several 
PWIA calculations~\cite{salme,rws93},
as well as a more recent and
more advanced calculation which fully includes
FSI~\cite{ishi98}. 
Thus, the inclusive asymmetry 
$A_{T'}$ in the vicinity of the $^{3}$He quasi-elastic peak
is most sensitive to the neutron magnetic form factor.


The experiment was carried out 
in Hall A at the Thomas Jefferson National Accelerator Facility (JLab),
using a longitudinally polarized continuous wave electron beam of
$10~\mu$A current incident on a high-pressure
polarized $^{3}$He gas target \cite{jlabtarget}.  The target was polarized by
spin-exchange optical pumping at a density of $2.5 \times 10^{20}$
nuclei/cm$^3$ using rubidium as the spin-exchange medium.
The beam and target polarizations were approximately 70\%
and 30\%, respectively, and the beam helicity was flipped at a
rate of 1 Hz (30 Hz for part of the experiment). To improve the optical
pumping efficiency, the target
contained a small admixture of nitrogen ($\sim$10$^{18}$ cm$^{-3}$). 
Backgrounds from the target
cell walls and the nitrogen admixture were determined in calibration
measurements using a reference cell with the same dimensions as those
of the $^3$He target cell.  The background levels were a few percent
of the full target yield. The background from rubidium was
negligible.  

Six kinematic points were measured corresponding to $Q^2 = 0.1$ to
$0.6$ (GeV/c)$^2$ in steps of 0.1 (GeV/c)$^2$.  An incident electron beam
energy of 0.778 GeV was employed for the two lowest $Q^2$ values of the 
experiment and the remaining points were completed at an incident beam 
energy of 1.727 GeV.
To maximize
the sensitivity to $A_{T'}$, the target spin was oriented at $62.5^\circ$
to the right of the incident electron momentum direction.
This corresponds to $\theta^\ast$ from $-8.5^\circ$ to $6^\circ$, 
resulting in a contribution to
the asymmetry due to $R_{TL'}$ of less than $2\%$ at all kinematic settings,
as determined from PWIA.  
To allow systematic checks, the target spin direction was rotated by
180$^\circ$ every 24-48 hours, and the overall sign of the beam
helicity was periodically reversed at the polarized electron injector
gun by inserting a $\lambda/2$ plate, resulting in four 
different combinations of beam and target polarization states.

Electrons scattered from the target were observed in the two Hall A
High Resolution Spectrometers, HRSe and HRSh. Both spectrometers were
configured to detect electrons in single-arm mode using nearly
identical detector packages consisting of two dual-plane vertical
drift chambers for tracking, two planes of segmented plastic
scintillators for trigger formation, and a CO$_2$ gas Cherenkov
detector and Pb-glass total-absorption shower counter for pion rejection.
  The spectrometer momentum and angular
acceptances were approximately $\pm$4.5\% and 5.5~msr, respectively.
The HRSe was set for quasi-elastic kinematics while the HRSh detected
elastically scattered electrons.  
Since the elastic asymmetry can
be calculated very well at low $Q^2$ using the well-known elastic form factors
of $^3$He~\cite{Amroun},
the elastic measurement allows precise monitoring of the product of the
beam and target polarizations, $P_tP_b$.
For the incident electron beam energy of 0.778 GeV, the HRSh was set
to $Q^2=0.1$ for the elastic scattering kinematics and $P_tP_b$ can be
determined to better than $2\%$. For the incident beam energy of 1.727
GeV, $P_tP_b$ can be determined to better than $3\%$ at $Q^2$ = 0.2 
(GeV/c)$^2$ for the elastic scattering.    
Standard M{\o}ller and NMR polarimetry were 
performed as a cross-check of the elastic polarimetry.
The $P_{t}P_{b}$ averaged over all six quasi-elastic kinematic
settings of this experiment determined from 
the elastic polarimetry was $0.208 \pm 0.001 \pm 0.005$, where the errors are
statistical and systematic, respectively. 
Combining the M{\o}ller and the NMR measurement, 
the average $P_{t}P_{b}$ was $0.215 \pm 0.013$ with the error being the
total systematic error.

\begin{figure}[t]
\psfig{file=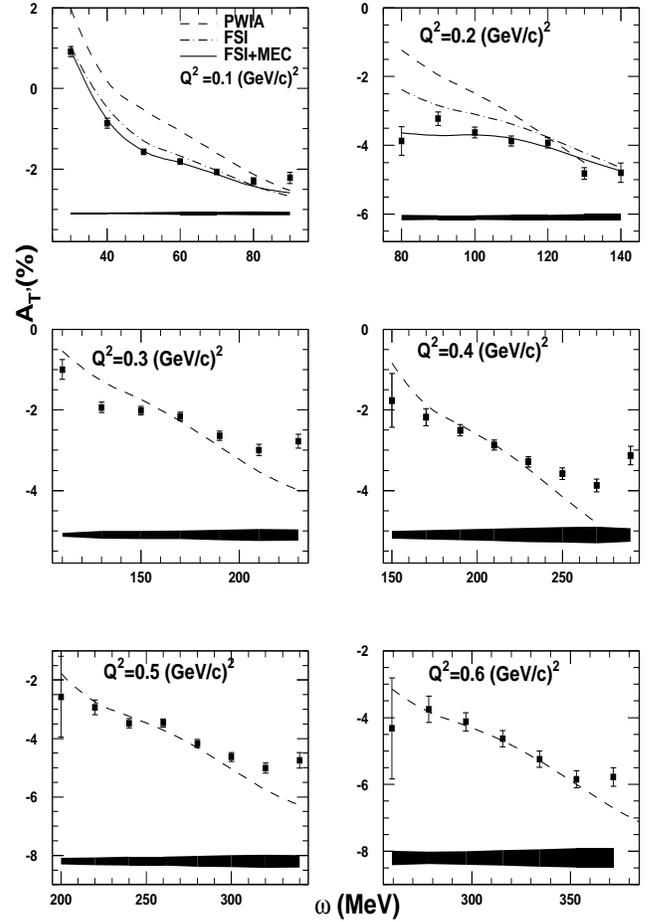,height=14.cm,width=9.cm}
\caption[]{The transverse asymmetry A$_{T'}$ at 
Q$^{2}$=0.1-0.6 (GeV/c)$^{2}$. The PWIA calculations are shown as dashed
curves. The Faddeev calculations which include FSI only and FSI and
MEC are shown as dash-dotted and solid curves, respectively.} 
\label{atq1}
\end{figure}


The yield for each electron helicity state was corrected by
its corresponding charge and computer deadtime, and 
the raw experimental asymmetry was extracted as a function of 
$\omega$ for all six kinematic settings. The raw asymmetry was
then corrected for dilutions due to scattering from the empty target walls, 
the nitrogen content and $P_{t}P_{b}$.
The physics asymmetry $A_{T'}$ 
was obtained after corrections for
radiative effects. 
Continuum radiative corrections were calculated using the covariant formalism
of Akushevich {\it et al.\/}~\cite{aku97}, which was generalized to 
quasi-elastic kinematics. This procedure requires knowledge of $^3$He nuclear 
response functions at various kinematic points.
These response functions were obtained from the full
Faddeev calculation for $Q^2 = 0.1$ and 0.2 (GeV/c)$^2$ and
the PWIA calculation~\cite{salme} for $Q^2 =0.3$ to 0.6 (GeV/c)$^2$.


Results for $A_{T'}$
 as a function of $\omega$ are shown in 
Fig.~\ref{atq1} for
all six kinematic settings of the experiment. The error bars on the data are
statistical only, and the total experimental systematic error is indicated
as an error band in each figure.
PWIA calculations~\cite{salme} using the AV18 for the NN interaction
potential and the H\"{o}hler nucleon form factor parametrization \cite{hoh}
are shown as dashed lines. 
The Faddeev calculations with FSI only  
and with both FSI and MEC using the Bonn-B potential and the
H\"{o}hler form factor parametrization 
are shown as dash-dotted lines 
and solid lines, respectively, for $Q^2 = 0.1$ and $0.2$ (GeV/c)$^2$.
All theory results were averaged over the spectrometer acceptances 
using a Monte Carlo simulation.
The systematic uncertainty in $A_{T'}$ includes contributions 
from $P_tP_b$, background subtraction, radiative corrections, 
helicity-correlated false asymmetries, 
and pion contamination.
A Monte Carlo simulation code was employed
to determine $P_{t}P_{b}$ from the measured
elastic asymmetry, taking into account
the spectrometer acceptance, energy loss, detector resolutions, 
and radiative effects. 
The total uncertainty in determining 
$P_{t}P_{b}$ is dominated by the uncertainties in the 
$^3$He elastic form factors.
The overall systematic uncertainty of 
$A_{T'}$ is 2$\%$ for $Q^2$ values of
0.1 and 0.2 (GeV/c)$^2$
dominated by the uncertainty in determining $P_{t}P_{b}$, 
and $5\%$ for $Q^2$ values of 0.3 to 0.6 (GeV/c)$^2$
dominated by the uncertainty in the radiative correction, which can be 
reduced with improved theoretical calculations for these values of $Q^2$.

\begin{figure}
\psfig{file=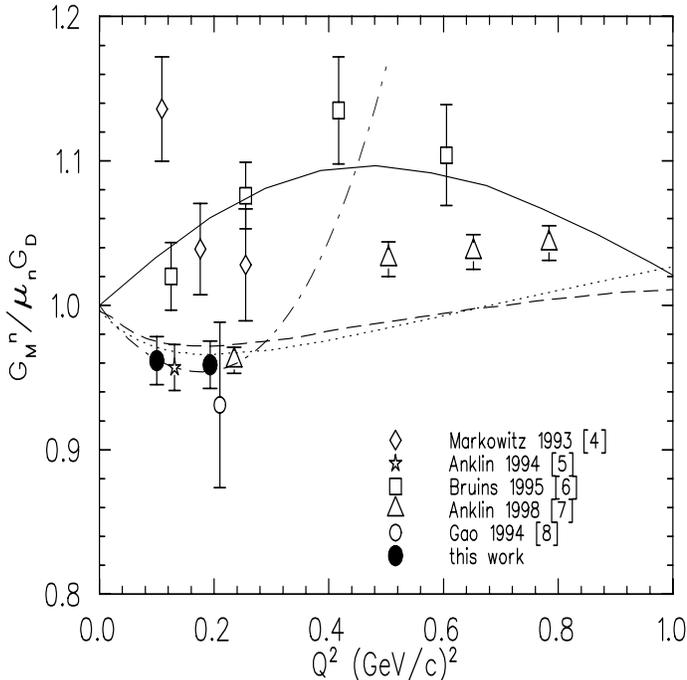,angle=90,height=9.cm,width=9.cm}
\caption[]{The neutron magnetic form factor $G_{M}^{n}$ in units of 
the standard dipole form factor $(1+Q^2/0.71)^{-2}$, as a function of
$Q^{2}$, along with previous measurements and theoretical models. The Q$^2$
points of Anklin 94~\cite{Ankl94} and Gao 94~\cite{Gao94} 
have been shifted 
slightly for clarity. The solid curve is a recent cloudy bag model 
calculation\cite{lu}, the long dashed curve is a recent calculation 
based on a fit of the proton data using dispersion 
theoretical arguments~\cite{mergell}, and 
the dotted curve is from the H\"{o}hler~\cite{hoh} parametrization. The  
dash-dotted curve is an analysis based on the relativistic baryon
chiral perturbation theory~\cite{kubis}.} 
\label{gmn}
\end{figure} 

The state-of-the-art three-body calculation 
treats the $^3$He target state and the 3N scattering states   
in the nuclear matrix element 
in a consistent way by solving the corresponding 3N Faddeev 
equations \cite{golak2}. 
The MEC effects were calculated using the prescription of 
Riska~\cite{riska85}, which includes $\pi$- and $\rho$-like exchange
terms.
While the agreement between the data and full calculations is very good
at Q$^{2}$ = 0.1 and 0.2 (GeV/c)$^{2}$, 
the full calculation is not expected to be applicable at higher
Q$^{2}$ because of its fully non-relativistic framework.
A full calculation within the framework of relativity is highly
desirable.   

To extract $G^n_M$ for the two lowest $Q^2$ 
kinematics, 
the transverse asymmetry data were averaged over
a 30 MeV bin around the quasi-elastic peak.  
The full Faddeev
calculation including MEC~\cite{golak} was employed to
generate $A_{T'}$ as a function of $G^n_M$ in the same $\omega$ region.
By comparing the measured asymmetries with the predictions, the $G^n_M$ values
at $Q^2=0.1$ and 0.2 (GeV/c)$^2$ were extracted. 
The extracted values of $G_{M}^{n}$ 
are shown in Fig.~\ref{gmn} along with results
from previous measurements and several theoretical calculations. 
The uncertainties shown are the quadrature sum of the statistical and 
experimental systematic uncertainties. These results are  
tabulated in Table ~\ref{gmn_tbl}. 

Since the full calculation described above 
is at present the only theoretical calculation 
available which treats FSI and MEC under
the present
experimental conditions, it is important to mention one highly nontrivial
internal test. The nuclear response functions for the 
inclusive scattering on $^3$He were
calculated in two independent ways by either integrating
explicitly over the pd and ppn break-up channels (with full inclusion
of FSI) or using a completeness relation \cite{ishikawa2}. 
The agreement between these two approaches is within $1\%$~\cite{golak3}. 
The Faddeev based formalism has been applied to other reaction
channels and good agreements have been found with experimental 
results \cite{golak3}, in particular the 
most recent NIKHEF data on $A^{0}_{y}$ at 
$Q^2 = 0.16$ (GeV/c)$^2$ from the
quasielastic $^{3}\vec{\rm {He}}(\vec{e},e'n)$ process \cite{hans}.

To investigate the theoretical uncertainty in extracting
$G^n_M$ at $Q^2$ =0.1 and 0.2 (GeV/c)$^2$, 
the full calculations were carried through with
two different NN potentials, Bonn-B and AV18. The difference in the
calculated asymmetries is less than $1 \%$ around the quasi-elastic peak.
The uncertainty 
due to $G^p_E$, $G^p_M$, and $G^n_E$ was studied by varying these
quantities over their experimental 
errors, and the range of variation 
in the calculated asymmetry was $1\%$.
The uncertainty due to MEC was estimated by comparing 
results with and without the inclusion of the $\Delta$ isobar current. 
At Q$^2$ = 0.1 and 0.2 (GeV/c)$^{2}$ relativistic corrections to $A_{T'}$
were estimated to be $2\%$ and $4\%$ \cite{billd} around the
quasi-elastic peak, respectively. 
Based on these studies, the overall theoretical 
uncertainty in calculating $A_{T'}$ was estimated to be $3.8 \%$ and
$5.1 \%$ for $Q^2 =0.1$ and 0.2 (GeV/c)$^2$, respectively.  
This results in an estimated theoretical uncertainty of $1.9\%$ 
and $2.6 \%$ in 
extracting $G^n_M$ for these two $Q^2$ points correspondingly, which
can be reduced once relativistic full calculations become available.
The errors on $G^n_M$ from present work shown in Fig.~2 and Table I are
experimental errors only, which do not include the theoretical
uncertainties discussed above.


In conclusion the inclusive transverse asymmetry $A_{T'}$
from the quasi-elastic $^3\vec{\rm He}(\vec{e},e')$ process
has been measured with
high precision at Q$^2$-values from 0.1 to 0.6 (GeV/c)$^2$. 
Using a full Faddeev calculation which
includes FSI and MEC we have extracted the neutron magnetic form factor
$G_{M}^{n}$ at Q$^2$ values of 0.1 and 0.2 (GeV/c)$^2$. 
The extracted values of $G_{M}^{n}$ at 
Q$^2$ of 0.1 and 
0.2 (GeV/c)$^2$ agree with the previous measurements of 
Anklin {\it et al.}~\cite{Ankl94,Ankl98}. 
The present experiment provides the first precision data on $G^n_M$ using
 a fundamentally different experimental approach than previous experiments.
 Thus it is a significant step towards understanding the
discrepancy among the existing data sets 
in the low-Q$^2$ region.
Although we have presented precise data on $A_{T'}$ at higher 
$Q^2$ (0.3 - 0.6 (GeV/c)$^2$) in this Letter, full  
calculations are at present not available for these values 
of $Q^2$ to allow the extraction of $G^n_M$ with high precision. 
Theoretical efforts are 
currently underway to extend the full calculation to higher Q$^{2}$ 
\cite{glockle}.


We thank the Hall A technical staff and the Jefferson Lab
Accelerator Division for their outstanding support during this experiment.
We also thank S.~Ishikawa for providing us with his 
calculations in the early stage of this work, T.~W.~Donnelly 
for helpful discussions, and O.N. Ozkul for the target polarization analysis.  
This work was supported in part by the U.~S.~Department of Energy, DOE/EPSCoR,
the U.~S.~National Science Foundation, 
the Science and Technology Cooperation
Germany-Poland and the Polish Committee for Scientific Research, 
the Ministero dell'Universit\`{a} e della Ricerca
Scientifica e Tecnologica (Murst),
the French Commissariat \`{a} l'\'{E}nergie Atomique,
Centre National de la
Recherche Scientifique (CNRS) and the Italian Istituto Nazionale di Fisica
Nucleare (INFN).
This work was supported by DOE contract DE-AC05-84ER40150
under which the Southeastern Universities Research Association
(SURA) operates the Thomas Jefferson National Accelerator Facility.
The numerical calculations were performed on the PVP machines at the U.~S. 
National Energy Research Scientific Computer Center (NERSC) and
the CRAY T90 of the NIC in
J\"{u}lich.


\begin{table}
\begin{tabular}{|c|c|c|} 
 $Q^{2}$ (GeV/c)$^{2}$ & $G_{M}^{n}/G_{M}^{n}(Dipole)$ &Uncertainties\\ \hline 
 0.1&0.962&$\pm$0.014$\pm$0.010\\ 
 0.2&0.959&$\pm$0.013$\pm$0.010\\
\end{tabular}
\caption[]{$G_{M}^{n}$ as a function of $Q^{2}$, the uncertainties are 
statistical and experimental systematic, respectively.} 
\label{gmn_tbl}
\end{table}

\end{document}